\newif\ifisonecolumn
\isonecolumntrue %use this to make it true (or comment if not)

\ifisonecolumn
\documentclass[onecolumn,draft,12pt]{IEEEtran}
\else
\documentclass{IEEEtran}
\fi

\input{defi_massive_ULDL.def}

\usepackage{amssymb}

\usepackage{etoolbox}
\newcommand{\FigDat}[2]{
\ifstrequal{#1}{Performance}{
\begin{figure}[t]
        \begin{center}\includegraphics[scale=#2,trim=8mm 0 0 3mm,clip]{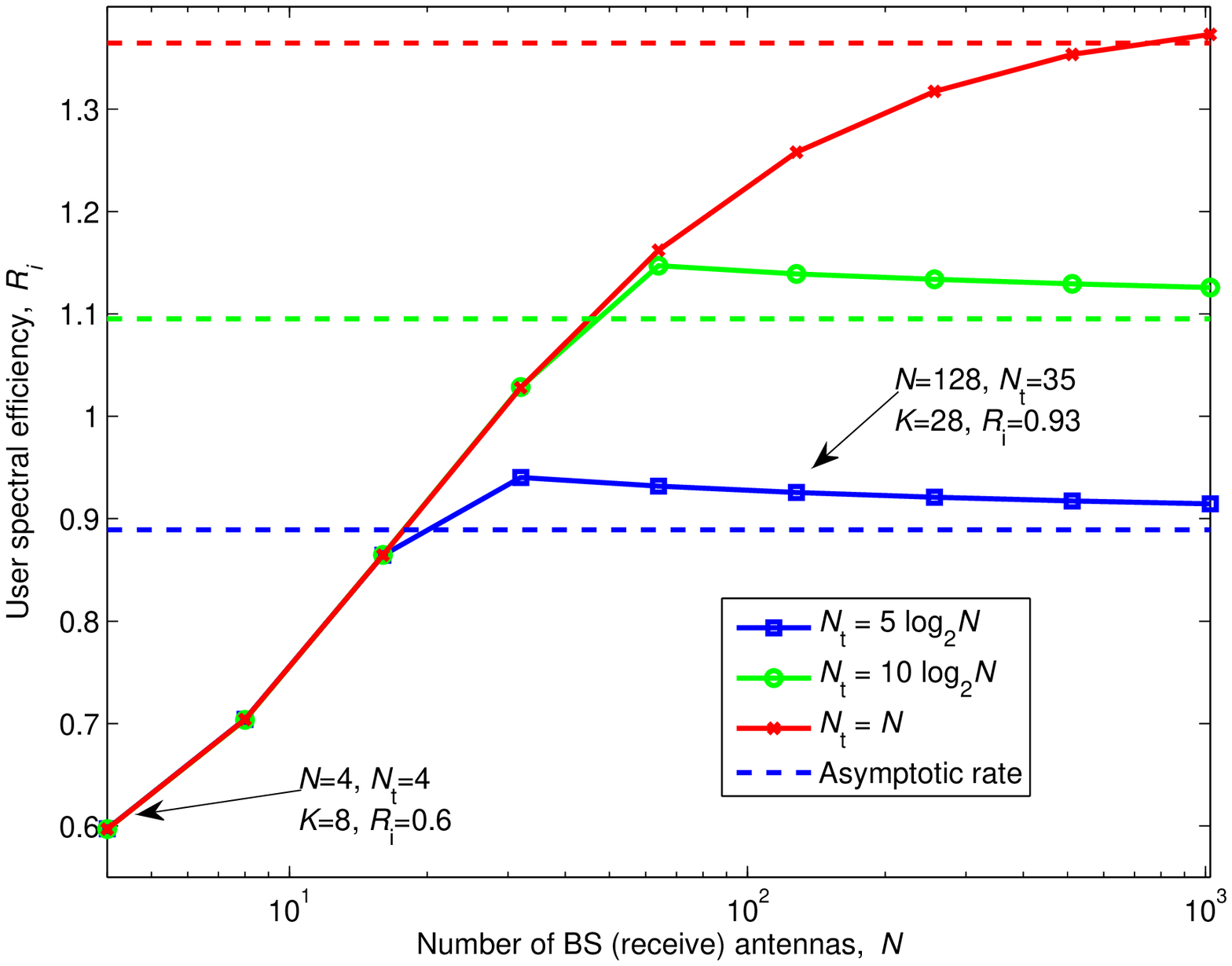}\end{center}
    \caption{Average spectral efficiency per user in the downlink of the massive MIMO system, as a function of the number of receive antennas ($c_\mathrm{u}=4$, $c_\mathrm{f}T=10$) THIS SHOULD APPEAR IN TEXT ALSO. The figure depicts the performance of a beamforming (BF) system, for various scalings of the number of antennas used for transmission.}
    \label{f:Performance fig}
\end{figure}
}{}
\ifstrequal{#1}{TotalPerformance}{
\begin{figure}[t]
        \begin{center}\includegraphics[scale=#2]{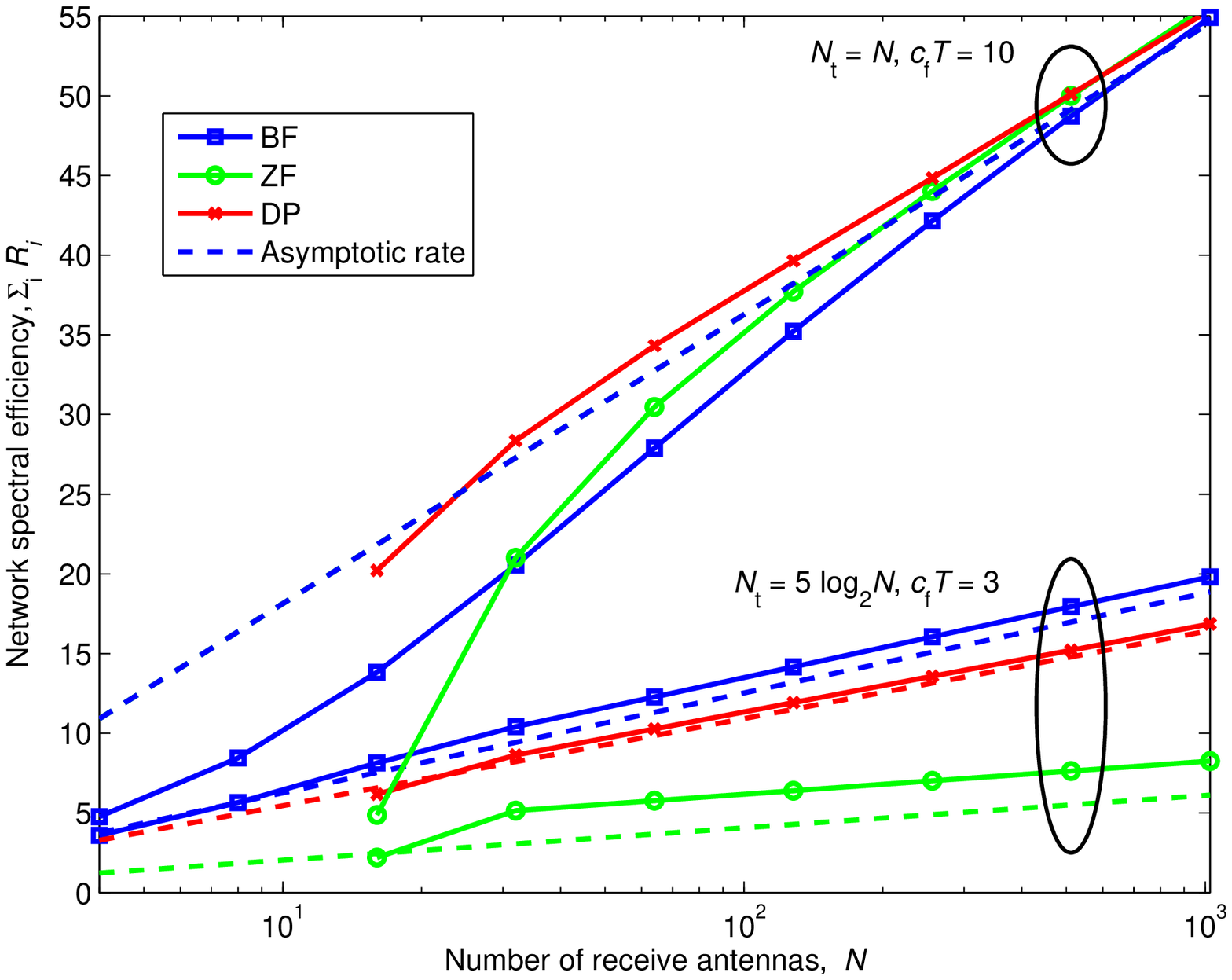}\end{center}
    \caption{Total spectral efficiency in the downlink of the massive MIMO system, as a function of the number of receive antennas. The number of served user is $K=c_\mathrm{u}\log_2 N$ with $c_\mathrm{u}=4$.}
    \label{f:TotalPerformance fig}
\end{figure}
}{}
\ifstrequal{#1}{Balancing}{
\begin{figure}[t]
        \begin{center}\includegraphics[scale=#2]{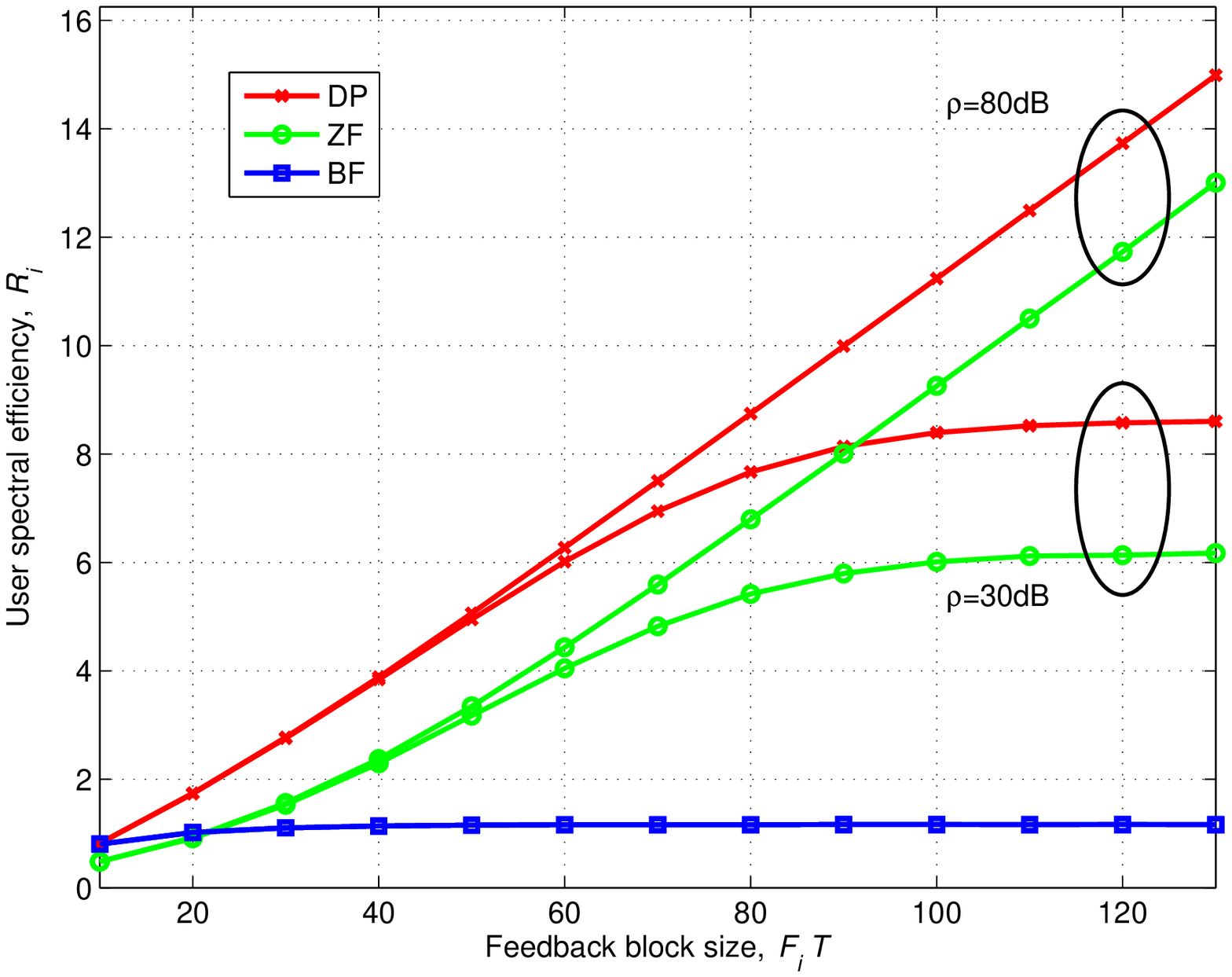}\end{center}
    \caption{Average spectral efficiency per user in the downlink of the massive MIMO system, as a function of the number of bits in feedback block. The system uses $N_\mathrm{t}=8$ transmit antennas and serves $K=8$ users.}
    \label{f:Balancing fig}
\end{figure}
}{}
}
\usepackage{amsbsy}

\begin{document}

\title{Uplink Downlink Rate Balancing and throughput scaling in FDD Massive MIMO Systems}
\author{Itsik~Bergel,~\IEEEmembership{Senior Member,~IEEE,}
        Yona~Perets,
        and~Shlomo~Shamai,~\IEEEmembership{Fellow,~IEEE}% <-this % stops a space
\thanks{\copyright {} 2015 IEEE. Personal use of this material is permitted. Permission from IEEE must be obtained for all other uses, in any current or future media, including reprinting/republishing this material for advertising or promotional purposes, creating new collective works, for resale or redistribution to servers or lists, or reuse of any copyrighted component of this work in other works.}% <-this % stops a space
\thanks{I. Bergel is with Faculty of Engineering, Bar-Ilan University, 52900 Ramat-Gan, Israel;
   (e-mail: bergeli@biu.ac.il).}% <-this % stops a space
\thanks{Y. Perets is with Marvell Semiconductor Israel, Azorim Park, Petach-Tikva, 49527, Israel; (e-mail: yoni@marvell.com).}% <-this % stops %a space
\thanks{S. Shamai (Shitz) is with the Department of Electrical Engineering, Technion-
Israel Institute of Technology, Technion City, Haifa 32000, Israel; (e-mail:
sshlomo@ee.technion.ac.il).}% <-this % stops a space
\thanks{Manuscript submitted July 9th, 2015.}}

\maketitle

\begin{abstract}
In this work we extend the concept of uplink-downlink rate balancing to frequency division duplex (FDD) massive MIMO systems. We consider a base station with large number antennas serving many single antenna users. We first show that any unused capacity in the uplink can be traded off for higher throughput in the downlink in a system that uses either dirty paper (DP) coding or linear zero-forcing (ZF) precoding. We then also study the scaling of the system throughput with the number of antennas in cases of linear Beamforming (BF) Precoding, ZF Precoding, and DP coding. We show that the downlink throughput is proportional to the logarithm of the number of antennas. While, this logarithmic scaling is lower than the linear scaling of the rate in the uplink, it can still bring significant throughput gains. For example, we demonstrate through analysis and simulation that increasing the number of antennas from $4$ to $128$ will increase the throughput by more than a factor of $5$. We also show that a logarithmic scaling of downlink throughput as a function of the number of receive antennas can be achieved even when the number of transmit antennas only increases logarithmically with the number of receive antennas.
\end{abstract}
%%%%%%%%%%%%%%%%%%%%%%%%%%%%%%%%%%%%%%%%%%%%
\section{Introduction}
In recent years, as cellular networks evolve, there has been increasing emphasis on the need for efficient transmission schemes. Next generation networks are required to deliver more throughput, e.g., increase their spectral efficiency, but at the same time also increase their energy efficiency.  
Massive MIMO, where each base station (BS) has many antennas, promises to bring a significant increase in the rates of cellular networks (e.g., \cite{rusek2013scaling}--\nocite{larsson2013massive,lu2013overview,de2013massive}\cite{zhengmassive} and references therein). It was also shown to improve the energy efficiency (EE), at least in some of the scenarios \cite{bjornson2013designing,chih2014toward}.

But, the realization of this capacity gain requires the availability of sufficiently accurate channel state information (CSI). Hence, much effort has been spent on the analysis and optimization of the training signals, channel estimation and CSI feedback (e.g., \cite{caire2010multiuser,kobayashi2011training}). Furthermore, many works have observed that the benefits of massive MIMO are limited in frequency division duplex (FDD) systems (e.g., \cite{rusek2013scaling}--\cite{de2013massive}) because of the difficulty to provide CSI to the base station. Thus, most of the analysis of massive MIMO system have focused on time division duplex (TDD) where the CSI can be measured by exploiting the channel reciprocity.

In practice, however, the majority of deployed systems employ frequency division duplex (FDD). Thus, although the gains are smaller than in TDD, it is still very important to be able to characterize the benefits of massive MIMO in FDD systems. In FDD systems, CSI is mostly obtained through feedback from the mobile terminals. Thus, much research has been devoted to the optimization of the CSI feedback in order to gain the most from the available antennas. 

One approach that has been considered is to exploit knowledge of the spatial correlation in the CSI feedback. Large gains were demonstrated using this approach (e.g., \cite{nam2012joint}--\nocite{choi2013noncoherent,choi2013downlink}\cite{kuo2012compressive}). The channel correlation changes quite slowly, and can be measured directly from the uplink transmissions. Hence, the knowledge of the channel correlation allows a significant reduction in the required feedback rate while achieving the same throughput. However, while the use of spatial correlation can increase the downlink rates, it cannot improve the asymptotic scaling of the rates with respect to the number of antennas. Hence, in this paper we will study the throughput scaling for the case of uncorrelated antennas, while keeping in mind that practical systems should also take advantage of the spatial correlation. 

As the CSI feedback requires uplink resources, the use of massive MIMO in FDD systems creates a tradeoff between the downlink throughput and uplink throughput. This tradeoff has been investigated in several works. However, the ability to use this tradeoff to dynamically control the balance between the uplink and downlink rates has been widely ignored. For example, 
 Caire et al. \cite{caire2010multiuser} and Kobayashi et al. \cite{kobayashi2011training} considered several types of CSI feedback, and the maximization of the downlink throughput for a given amount of overhead (channel uses for training and feedback). But, in practical systems the uplink load and the downlink load can differ significantly.  Hence, the `price' of uplink resources is different for different systems, and may also change with time. We also note that the systems in  \cite{caire2010multiuser,kobayashi2011training} did not specifically consider massive MIMO, and do not scale well with the number of antennas. In particular, both works assume that the number of users is equal to or larger than the number of antennas. Such a scenario allows a linear growth of the throughput with the number of antennas only if the feedback rate also scales linearly with the number of antennas (see also \cite{jindal2006mimo}). But, as will be demonstrated in Section \ref{sub: proof blancing}, the uplink cannot support such a large amount of feedback. 

In contrast to the current literature, our focus here is on the scaling of the throughput with the number of antennas in the downlink of FDD massive MIMO systems.
In the following we show that the downlink sum-rate in an FDD system can scale logarithmically with the number of antennas. This is due to the fact that downlink throughput is linearly related to the feedback rate (e.g., \cite{caire2010multiuser,kobayashi2011training,jindal2006mimo}), but the uplink can only support an increase in feedback rate that is logarithmic in the number of base station antennas.

The logarithmic  scaling of downlink throughput in FDD systems is in contrast to the linear scaling achievable in TDD systems, where channel reciprocity can be used for downlink channel estimation and hence there is no need for CSI feedback. For example, it is shown in \cite{rusek2013scaling,lu2013overview} that the number of served TDD users scales linearly with the number of antennas, and that the per-user rate converges to a constant; thus, the total system throughput scaled linearly. Note that although the asymptotic user rate depends on various factors (e.g., scheduling, channel estimation, pilot contamination, processing type etc.), the throughput scaling  remains linear as long as the channel coherence block (in time and frequency) is large enough. Some works have also addressed the system optimization with `relatively small' number of antennas, and even showed that such optimization can significantly reduce the number of antennas that are required to achieve a specific throughput gain \cite{huh2012achieving,hoydis2013massive}. However, even in these cases, the asymptotic scaling remains linear.

In this work we adopt the uplink-downlink balancing principle introduced in \cite{bergel2012IEEEI,yellin2011balancing,bergel2014balancing_submitted}, and use it to characterize the throughput scaling of both link directions in an FDD system as the number of antennas grows.
The uplink-downlink balancing principle suggests the use of available uplink capacity to increase the downlink capacity. The relevance to massive MIMO systems is obvious: the uplink enjoys the most gain due to the availability of CSI at the BS. Some of this uplink capacity increase can be traded off to also increase the downlink throughput. 

We first demonstrate that any unused uplink rate can be converted to an increase in the downlink rate with a favorable exchange ratio. This result can be proven trivially for systems that perform dirty paper coding based on the results in  \cite{bergel2012IEEEI,bergel2014balancing_submitted}. We also provide a novel proof that shows that the same exchange ratio is achievable in systems that perform linear zero-forcing precoding.  

Subsequently, we also consider a scenario in which a constant fraction of the uplink rate is devoted to CSI feedback, and show that the downlink sum-rate can scale logarithmically with the number of antennas. Note that this result differs from the results in  \cite{bergel2012IEEEI,bergel2014balancing_submitted}, as
the scaling here is with respect to the number of antennas, and not with respect to the signal
to noise ratio (SNR). The logarithmic scaling of the downlink rate is lower than the linear uplink-rate or TDD-rate scaling, but still provides good throughput gains, which we further illustrate with simulation results. For example, increasing the number of antennas from $4$ to $128$ can increase the throughput by a factor of $5$. Furthermore, we also show that this logarithmic scaling on downlink throughput can be achieved even if the number of transmit antennas grows only logarithmically with the number of base station receive antennas. This important result helps minimize the pilot reference symbols required on the downlink and lowers implementation cost and complexity.

%%%%%%%%%%%%%%%%%%%%%%%%%%%%%%%%%%%%%%%%%%%%
\section{System model}\label{sec: system model}
We consider a cellular system with $K$ single antenna mobiles, and one BS with $N$ antennas. Assuming a flat fading channel and focusing on the downlink, the symbols that are received by all the mobiles at a given time can be written in a vector form as:
\begin{IEEEeqnarray}{rCl}\label{d: channel model}
\yv = \sqrt{\rho}\Hm\xv +\nv
\end{IEEEeqnarray}
where $\Hm$ is the $K \times N_\mathrm{t}$ complex valued channel matrix, whose $i,j$-th element, $\ch_{i,j}$,  is the channel gain from the $i$-th BS to the $j$-th mobile, $\xv$ is an $N_\mathrm{t}  \times 1$ complex vector of
the transmitted symbols, and $\nv$ is an $K \times 1$ noise vector whose elements are mutually independent proper complex Gaussian random variables \cite{neeser1993proper}
with zero mean and unit variance. We assume that all BSs antennas are used for reception, but not necessarily for transmission, i.e., we 
 consider the general case where the number of transmit antennas may be less than the number of receive antennas,  $N_\mathrm{t}\le N$. For normalization, we assume that the power of each channel element is $E[|\ch_{i,j}|^2]=1$, and $\rho$ denotes the SNR.
The transmitted symbols must satisfy the sum power constraint: 
\begin{IEEEeqnarray}{rCl}\label{e: x power constraint}
E[\|\xv\|^2] \le 1 .
\end{IEEEeqnarray}

We assume a block fading channel model, so that $\Hm$, is assumed to be fixed during the transmission of a block of $T$ symbols (in different times or frequencies), but changes in an independent manner from
block to block. The channel elements are further assumed to be mutually independent, proper, complex Gaussian random variables with zero mean.

In this paper we focus on the case of perfect channel estimation at the mobile receivers, and
imperfect (i.e., quantized) CSI at the BS transmitter.
Thus, we assume that the receivers can measure the channels without errors. Note that the effects of channel estimation, pilot signals 
and pilot contamination can be very significant in massive MIMO systems (e.g., \cite{jose2011pilot}--\nocite{hoydis2011massive,yin2012coordinated}\cite{lozano2013fundamental}), particularly for the case of short coherence time. Yet, as most cellular FDD systems have relatively large coherence times, this paper focuses on the channel feedback, and leaves the affects of channel estimation for future study\footnote{It is also important to note that the presented rate scaling result is achievable when the number of transmit antennas scales logarithmically with the number of receive antennas. Thus, the required number of pilot symbol will typically be quite low compared to the channel coherence time, and the challenges of channel estimation are not very large. }.

%%%%%%%%%%%%%%%%%%%%%%%%%%%%%%%%%%%%%%%%%%%%
\section{Uplink-Downlink rate tradeoff}\label{sec: uplink downlink tradeoff}
Our main results characterize the scaling of an FDD massive-MIMO system with the number of antennas. We show that the number of served users in the downlink can scale logarithmically with the number of antennas. 
This scaling is obtained by using part of the capacity gain of the uplink (due to the use of massive-MIMO) to send a large amount of feedback in the downlink, using the uplink-downlink balancing concept. These results give motivation for the utilization of massive-MIMO in FDD systems, and not only in TDD systems.

We start by showing that any un-needed rate in the uplink of each user can be traded into a downlink rate, with a good exchange ratio.
We define $F_i$ as the total feedback rate sent by mobile $i$ (over the uplink) and $R_i$ as the downlink rate achievable by mobile $i$. 
\begin{theorem}[Massive-MIMO uplink downlink balancing]\label{Th: DP balancing} In a system that uses either dirty paper coding (DP) or linear zero-forcing precoding (ZF), the ratio between the downlink rate  and the corresponding feedback rate at the interference limited regime is lower bounded by:
\begin{IEEEeqnarray}{rCl}\label{e: th1_bound}
\lim_{F_i\rightarrow \infty}\lim\limits_{\rho \rightarrow \infty}\frac{R_i}{F_i}\ge\frac{T}{N_\mathrm{t}}
\end{IEEEeqnarray}
\end{theorem}
\begin{IEEEproof}The proof for a system that applies DP is a straight forward adaptation of Theorem 1 in \cite{bergel2012IEEEI,bergel2014balancing_submitted} to the case of Massive MIMO. The proof for a system that applies ZF is given in Subsection \ref{sub: proof blancing}.
\end{IEEEproof}

The exchange ratio of Theorem \ref{Th: DP balancing} ($T/N_\mathrm{t}$) is favorable as long as the size of a channel coherence block (in symbols) is larger than the number  of transmit antennas. This is good news because in many cellular scenarios the coherence block includes hundreds of symbols or more (see \cite{bergel2014balancing_submitted} for an example). 

Interestingly, the theorem holds for each mobile separately, i.e., the uplink-downlink ratio can be balanced for each mobile separately according to its needs and capabilities. Nevertheless, in the rest of the paper we focus on the case where all users have identical service requirements and capabilities, and study the scaling of the number of supported users with the number of antennas.

Theorem \ref{Th: DP balancing} describes the uplink downlink tradeoff for a given number of transmit antennas. We next turn to study the scaling as the number of antennas grows to infinity. As the uplink user rate scales only logarithmically with the number of antennas, the available feedback is not sufficient to allow linear scaling of the number of users with the number of antennas\footnote{For example \cite{caire2010multiuser} (using linear ZF) and \cite{bergel2012IEEEI} (using DP) required a feedback rate per user that grows linearly with the number of transmit antennas.}. In the following we show that using either a DP scheme, a linear beamforming scheme or a linear ZF scheme can allow a logarithmic scaling of the number of users with the number of antennas, while keeping a constant rate per user. 
This scaling is summarized by the following theorem: 

\begin{theorem}[Linear precoding]
\label{Th: Scaling} At the massive MIMO asymptote, devoting a fraction, $c_{\rm f}$, of the uplink rate of each mobile to feedback, for any $c_\mathrm{u}>0$, a system that employs linear precoding or DP can support $K=c_\mathrm{u}\log_2 N$ mobiles while providing a constant rate for each mobile that is lower bounded by: 
\begin{IEEEeqnarray}{rCl}\label{e: final uesr rate total}
\lim_{N \rightarrow \infty} R_i&\ge&
\log_2\left(1+\frac{\rho }{1+\rho }\cdot\frac{ c_\mathrm{f} T}{c_\mathrm{u}\log_2 e }\right).
\end{IEEEeqnarray}
The same number of users can also be supported with only $N_\mathrm{t}=c_\mathrm{t} \log_2 N$ transmit antennas, for any $c_\mathrm{t}>0$ in the case of linear beam-forming precoding (BF) and for any $c_\mathrm{t}\ge c_\mathrm{u}$ in the case of  ZF or DP. In such case, the rate of each mobile can be bounded by: 
\begin{IEEEeqnarray}{rCl}\label{e: uesr rate limit ct}
\lim_{N \rightarrow \infty} R_i
&\ge&
\log_2\left(1+\rho\frac{ \left(\frac{c_{\rm t}}{c_{\rm u}}-\beta\right)\left( 1-2^{-\frac{c_{\rm f}T }{c_{\rm t} }} \right)}{ 1+\rho(1-\beta) +\rho    \beta2^{-\frac{c_{\rm f}T }{c_{\rm t} }}  }\right)+\Delta \IEEEeqnarraynumspace
\end{IEEEeqnarray}
where $\beta=0$ for BF and $\beta=1$ for ZF or DP, while $\Delta=0$ for BF and ZF, and 
\begin{IEEEeqnarray}{rCl}\label{d: Delta DP}
\Delta= -\frac{ (1 + \frac{c_\mathrm{t}}{c_{\rm u}}\gamma )  \log_2\left(1-\frac{\gamma}{1 +\frac{c_\mathrm{t}}{c_{\rm u}} \gamma }\right)}{\gamma }
-\log_2(e)>0 
\end{IEEEeqnarray}
for DP, with
\begin{IEEEeqnarray}{rCl}
\gamma &=&\rho\frac{1-2^{-\frac{ c_\mathrm{f}  T }{ c_\mathrm{t}  }}}{1+\rho  2^{-\frac{ c_\mathrm{f}  T }{ c_\mathrm{t}  }}}.
\end{IEEEeqnarray} 
\end{theorem}
\begin{IEEEproof} See Section \ref{sub: scaling proof}.
\end{IEEEproof}
The bounds of Theorem \ref{Th: Scaling}
are appealing because they show a significant rate increase in the downlink of an FDD system through the use of massive MIMO. Furthermore, the ability to achieve this scaling while the number of transmit antennas scale only logarithmically with the number of receive antennas (even while only using beam-forming) significantly  reduces the implementation complexity. Note that
the
large number of receive antennas is still needed, in order  to increase the uplink capacity and allow the required scaling of the feedback rate.

Note that no 
scheme is uniformly best. 
As $\Delta$ is monotonic increasing with $\frac{c_\mathrm{u}}{c_{\rm t}}$, we see that $\Delta$ is lower bounded by $0$ and hence, DP is (asymptotically) always better than ZF\footnote{Recalling that for DP and ZF $c_\mathrm{u}\le c_{\rm t}$, and also noting that $\gamma<1$, we also have that $\Delta\le \frac{ (1 + \gamma )  }{\gamma }\log_2(1 + \frac{\gamma}{1 +\gamma } )-\log_2(e)<2-\log_2(e)=0.56$, and hence the difference between the two schemes is bounded.}.  
The relation between the DP and the BF schemes is more difficult to characterize. However, one can easily verify that the BF scheme is asymptotically preferable over the ZF scheme if and only if:
\begin{IEEEeqnarray}{rCl}\label{e: ZF condition}
\frac{c_\mathrm{u}}{c_\mathrm{t}}>(1-2^{-\frac{c_\mathrm{f} T}{c_{\rm t} }}) \frac{\rho}{1+\rho}.
\end{IEEEeqnarray}
Since the DP scheme outperforms the ZF scheme, Condition (\ref{e: ZF condition}) is not sufficient to show that BF is preferable to DP. Yet, an inspection of (\ref{e: uesr rate limit ct}) shows (similarly to (\ref{e: ZF condition})) that BF is preferable over DP if the SNR, $\rho$, or the feedback fraction $c_\mathrm{f}$  are small enough, or if the number of users is large enough (recall that the BF scheme can also work when the number of users is larger than the number of transmit antennas, i.e., $c_\mathrm{u}>c_\mathrm{t})$.

%%%%%%%%%%%%%%%%%%%%%%%%%%%%%%%%%%%%
\section{Proof of Theorems \ref{Th: DP balancing} and \ref{Th: Scaling}}\label{sec: proof}
The proof assumes high rate feedback from the mobiles to the BS and precoding of the transmission to each mobile using this feedback. The structure of the suggested system is described in the following Subsection, leading to the proof of Theorem \ref{Th: DP balancing}. The proof of Theorem \ref{Th: Scaling} is given in Subsection \ref{sub: scaling proof}.
\subsection{Proof of Theorem \ref{Th: DP balancing}}\label{sub: proof blancing}
\subsubsection{Uplink rate}\label{Subsub: uplink}
As the first step, we characterize the uplink capacity, in order to evaluate the available feedback rate.
In an FDD scenario, the uplink gains the largest improvement, as the BS can directly estimate the uplink channels. Let $\Hm_{\rm U}$ denote the uplink $N \times K$ channel matrix, the uplink received signal is given by: 
\begin{IEEEeqnarray}{rCl}
\yv_\U=\sqrt{\rho}\Hm_\U \xv_\U+\nv_\U.
\end{IEEEeqnarray}
The achievable user rates 
in the multiple access channel were derived in \cite{verdu1999spectral} for various detectors.  For simplicity, we present here only the analysis of the linear MMSE receiver (which is obviously a lower bound on the uplink capacity). This MMSE bound is given by:
\begin{IEEEeqnarray}{rCl}\label{e: Rul lower bound}
R_{\U,i} 
\TwoOneColumnAlternate{\hspace{-1mm}=\hspace{-1mm}}{=}
E\left[\log_2\left(1+\rho\hv_{\U,i}^H\left(\rho\Hm_{\U,\bar i} \Hm_{\U,\bar i}^H+I\right)^{-1}\hv_{\U,i}\right)\right].\IEEEeqnarraynumspace
\end{IEEEeqnarray}
where $\hv_{\U,i}$ and $\Hm_{\U,\bar i}$ denote the $i$-th column of $\Hm_{\U}$ and the remaining matrix with all other columns of $\Hm_{\U}$, respectively. 

As $N$ goes to infinity, and if $\lim_{N\rightarrow\infty}K/N<1$, then \cite{CIT-001}:
\begin{IEEEeqnarray}{rCl}\label{e: MMSE UL bound}
\lim_{N\rightarrow \infty}\frac{\rho\hv_{\U,i}^H\left(\rho\Hm_{\U,\bar i} \Hm_{\U,\bar i}^H+I\right)^{-1}\hv_{\U,i}}{N}
\TwoOneColumnAlternate{\hspace{-1mm}=\hspace{-1mm}}{=}
\rho(1-\lim_{N\rightarrow\infty}\frac{K}{N})\IEEEeqnarraynumspace
\end{IEEEeqnarray}
and hence:
\begin{IEEEeqnarray}{rCl}\label{e: Rul scaling}
\lim_{N\rightarrow\infty}\frac{R_{\U,i}}{\log_2 N} \ge 1.
\end{IEEEeqnarray}
Thus the uplink can support  any number of mobiles which grows at most linearly with $N$ with a rate that grows logarithmically for each mobile. (In the following we choose the number of users, $K$, to grow only logarithmically with $N$, and hence the limit in (\ref{e: MMSE UL bound}) satisfies $\lim_{N\rightarrow\infty}\frac{K}{N}=0$.) 
\subsubsection{Feedback quantization}
As mentioned above, we assume that each mobile has perfect knowledge of the channels. Each mobile quantizes its measured channels and sends the index of the quantized version to the BS. We write the channel matrix as the sum:
\begin{IEEEeqnarray}{rCl}\label{d: quantization error representation}
\Hm =\hHm +\eM
\end{IEEEeqnarray}
where $\hHm$ is the quantized channel known to the transmitter, while $\eM$ is the channel error that is not known to the transmitter. For simplicity we assume an optimal, multi-dimensional, quantization scheme\footnote{See \cite{bergel2014balancing_submitted} for an analysis of a simpler quantization scheme that can also be adapted to the problem discussed here.}.  Using rate distortion theory, the rate that is required to quantize the channel gains of the $N_\mathrm{t}$ antennas to a quantization mean square error (MSE) of $E[|\epsilon_{i,j}|^2]=\XMe^2 _{i}$ is \cite{berger1971rate,cover2006elements}:
\begin{IEEEeqnarray}{rCl}\label{e: optimal feedback rate formula}
 F_{i}=\begin{cases}\frac{N_\mathrm{t}}{T}
\log_2\left(\frac{1 }{\XMe^2 _{i}}\right) & \XMe^2 _{i}<1
\\ 0 & \XMe^2 _{i}=1
\end{cases}
\end{IEEEeqnarray}
where we have assumed that the feedback is sent only once for a coherence block of  $T$ symbols.
Furthermore, by the asymptotic properties of the rate distortion function (e.g., \cite{cover2006elements} chapter 10.3), each quantization error element, $\epsilon_{i,j}$, is a proper complex Gaussian variable with zero mean, which is statistically independent of the quantized channel, $\hHm$, the transmitted symbols, $\xv$
and every other element of $\eM$.
Thus, we have:
\begin{IEEEeqnarray}{rCl}
E\left[|\hat h _{i,j}|^2\right] = 1-\XMe_i^2.
\end{IEEEeqnarray}

\subsubsection{Linear precoding}
%\subsection{Transmit precoding}
The transmitted vector is generated by the linear precoding: \begin{IEEEeqnarray}{rCl}
\xv=\Um^H \sv
\end{IEEEeqnarray}
where $\Um$ is precoding matrix and  $\sv$ is the $K \times 1$ vector that contains the actual data symbols, which are assumed to be iid proper complex Gaussian  random variables with zero mean and unit variance. 

To derive a general result that includes both the BF and the ZF schemes as special cases, we consider a scheme that performs partial ZF, i.e., tries to zero the interference to $L$ users, where $0\le L \le K-1$. All other degrees of freedom are used to increase the received signals' powers. Thus, each row of $\Um$ is generated from the corresponding row in $\hHm$ by first zero-forcing the signal to $L$ users and then normalizing its power. In mathematical terms, denoting by $\hv_{i}$ the  $i$-th row of $\hHm$ and by $\Pm_i$ the projection matrix to the null space of the $L$ selected interfered users, we define $\thv_{i}=
 \hv_{i} \Pm_i$, and  the $i$-th row of  $\Um$ is given by:
\begin{IEEEeqnarray}{rCl}
\uv_{i}=\frac{\thv_{i}}{\sqrt{K}\sqrt{\thv_{i} \thv_{i}^H}}.
\end{IEEEeqnarray}
Note that the $\sqrt{K}$ term in the denominator satisfies the power constraint, (\ref{e: x power constraint}), using $E[\sv\sv^H]=\mathbf{I}$.

The resulting effective channel is:
\begin{IEEEeqnarray}{rCl}\label{e: revised channel model}
\yv =  \sqrt{\rho}\Hm \Um^H\sv +\nv.
\end{IEEEeqnarray}
Thus, the achievable rate of user $i$ is lower bounded by:
\begin{IEEEeqnarray}{rCl}\label{e: uesr rate total}
R_i
&\ge&
E\left[
\log_2\left(1+\frac{\rho|\hv_{i}\uv_{i}^H |^2 E[|s_i|^2]}{\rho  \sum_{k\ne i}  |\hv_{i}\uv_{k}^H |^2 E[|s_k|^2]+1}\right)
\right] 
.
\IEEEeqnarraynumspace\end{IEEEeqnarray}
\subsubsection{Rate balancing}

In order to study the rate balancing tradeoff, we need to characterize the distribution of the numerator and  the denominator of (\ref{e: uesr rate total}). 
For the random term in the numerator we use:
\begin{IEEEeqnarray}{rCl}\label{e: nom statistics}
\hv_{i}\uv_{i}^H &=&\hhv_{i}\uv_{i}^H +\eM_{i}\uv_{i}^H 
\nonumber \\
&=&\frac{\hhv_{i}\Pm_i \hhv_{i}^H}{\sqrt{K}\sqrt{\hhv_{i}\Pm_i \hhv_{i}^H}} +\eM_{i}\uv_{i}^H 
\nonumber \\
&=&\frac{\sqrt{\hhv_{i}\hhv_{i}^H}}{\sqrt{K}}\frac{\sqrt{\hhv_{i}\Pm_i \hhv_{i}^H}}{\sqrt{\hhv_{i}\hhv_{i}^H}} +\eM_{i}\uv_{i}^H 
\nonumber \\
&=&\frac{\sqrt{1-\XMe_i^2}}{\sqrt{2 K}}r_i v_i +\frac{\XMe_i}{\sqrt{K}} w_i \end{IEEEeqnarray}
where $r_i^2$ has a chi-square distribution with $2 N_{\rm t}$ degrees of freedom, $v_i$ has a beta distribution, $v_i\sim\beta(N_{\rm t}-L,L)$, and $w_i$ has a complex normal distribution with zero mean and a variance of $1$.
For the denominator we have:
\begin{IEEEeqnarray}{rCl}\label{e: denom statistics}
  \sum_{k\ne i}  |\hv_{i}\uv_{k}^H |^2 &=&  \sum_{k\ne i,k\notin\mathcal{L}_i}  |\hv_{i}\uv_{k}^H |^2 +\sum_{k\ne i,k\in\mathcal{L}_i}  |\eM_{i}\uv_{k}^H |^2 
 \nonumber \\&=&\frac{1}{2 K}(p_i^2+\XMe_i^2g_i^2)
\end{IEEEeqnarray}
where $p_i^2$ has a chi-square distribution with $2 (K-1-L)$ degrees of freedom and $g_i^2$ has a chi-square distribution with $2 L$ degrees of freedom, and $\mathcal{L}_i$ is the set of users that perform interference cancellation for user $i$. 

Using (\ref{e: optimal feedback rate formula}), we have $\XMe_i^2=2^{-\frac{F_i T}{N_{\rm t} }}$. Substituting $\XMe_i^2$ together with (\ref{e: nom statistics}), (\ref{e: denom statistics}) and $E[|s_i|^2]=1$ into (\ref{e: uesr rate total}) gives:
\begin{IEEEeqnarray}{rCl}\label{e: uesr rate total 4}
R_i
\TwoOneColumnAlternate{&\hspace{-1mm}\ge\hspace{-1mm}&}{&\ge&}
E\left[
\log_2\left(1+\frac{\rho\left|\frac{\sqrt{1-2^{-F_i T/N_{\rm t} }}}{\sqrt{2 K}}r_i v_i +\frac{\sqrt{2^{-F_i T/N_{\rm t} }}}{\sqrt{K}} w_i \right|^2 }{\frac{\rho}{2 K}(p_i^2+2^{-\frac{F_i T}{N_{\rm t} }} g_i^2)+1}\right)
\right]  
\TwoOneColumnAlternate{\nonumber \\}{}
\end{IEEEeqnarray}
and taking the limit as $\rho$ grows to infinity gives the interference limited rate:
\begin{IEEEeqnarray}{rCl}\label{e: uesr rate total 5}
\TwoOneColumnAlternate{&&}{}
\lim_{\rho\rightarrow \infty}R_i
\TwoOneColumnAlternate{\\ \nonumber&& \ge}{&\ge&}
E\left[
\log_2\left(1+\frac{\left|\sqrt{1-2^{-F_i T/N_{\rm t} }}r_i v_i +\sqrt{2^{1-F_i T/N_{\rm t} }} w_i \right|^2 }{p_i^2+2^{-\frac{F_i T}{N_{\rm t} }} g_i^2}\right)
\right]  
\end{IEEEeqnarray}
If $L<K-1$ then the limit of (\ref{e: uesr rate total 5}) is bounded as $F_i$ grows to infinity. On the other hand, in the ZF case $L=K-1$, and we get $p_i=0$ and:
\begin{IEEEeqnarray}{rCl}\label{e: uesr rate total 6}
\lim_{F_i\rightarrow \infty}\lim_{\rho\rightarrow \infty}\frac{R_i}{F_i}
&\ge&\lim_{F_i\rightarrow \infty}\frac{1}{F_i}
E\left[
\log_2\left(1+\frac{\left|r_i v_i  \right|^2 }{2^{-\frac{F_i T}{N_{\rm t} }} g_i^2}\right)
\right]
\nonumber \\
&=&\lim_{F_i\rightarrow \infty}\frac{1}{F_i}
\left(\frac{F_i T}{N_{\rm t} }+E\left[
\log_2\left(\frac{\left|r_i v_i  \right|^2 }{ g_i^2}\right)
\right]\right)
\nonumber \\
&=&
\frac{T}{N_{\rm t} }
\end{IEEEeqnarray}
which completes the proof of Theorem \ref{Th: DP balancing}.
\subsection{Proof of Theorem \ref{Th: Scaling}}\label{sub: scaling proof}
\subsubsection{Proof for linear precoding}\label{subsub: linear precoding scaling}
We begin by proving Theorem \ref{Th: Scaling} for the BF and ZF linear precoding schemes. We rewrite (\ref{e: uesr rate total}) as:
\begin{IEEEeqnarray}{rCl}\label{e: uesr rate total for balancing}
R_i
\TwoOneColumnAlternate{&\hspace{-1mm}\ge\hspace{-1mm}&}{&\ge&}
E\left[
\log_2\left(1+\frac{\rho\frac{K}{N_{\rm t}-L}\frac{|\hv_{i}\uv_{i}^H |^2}{(1-\XMe_i^2)}}{\frac{N_{\rm t}}{N_{\rm t}-L}\frac{K}{(1-\XMe_i^2)N_{\rm t} }\left(\rho  \sum_{k\ne i}  |\hv_{i}\uv_{k}^H |^2 +1\right)}\right)
\right] 
\TwoOneColumnAlternate{\nonumber \\}{}
\end{IEEEeqnarray}
and set the feedback rate to be $F_i=c_\mathrm{f} \log_2 N$. Considering the different terms in (\ref{e: uesr rate total for balancing}), we use the following limits: 
\begin{IEEEeqnarray}{rCl}\label{e: S bound total}
\TwoOneColumnAlternate{&&}{}
\lim_{N_{\rm t}-L\rightarrow\infty}\frac{K}{N_{\rm t}-L}\frac{|\hv_{i}\uv_{i}^H |^2}{(1-\XMe_i^2)} 
\TwoOneColumnAlternate{\nonumber \\ && \hspace{1cm}=}{=}
\lim_{N_{\rm t}-L\rightarrow\infty}\frac{|\hhv_{i}\thv_{i}^H +\eM_{i}\thv_{i}^H|^2 }{(N_{\rm t}-L) (1-\XMe_i^2)\thv_{i} \thv_{i}^H}= 1
\end{IEEEeqnarray}
\begin{IEEEeqnarray}{rCl}\label{e: I bound total}
\lim_{K\rightarrow\infty}\sum_{k\ne i}  |\hv_{i}\uv_{k}^H |^2
&=&\lim_{N_{\rm t}\rightarrow\infty}\frac{L \XMe_i^2+K-L}{K}
\nonumber \\
&=&1-(1-\xi^2)\beta.
\end{IEEEeqnarray}
and
\begin{IEEEeqnarray}{rCl}\label{e: limit Nt L}
\lim_{N_{\rm t}\rightarrow\infty}\frac{N_{\rm t}}{N_{\rm t}-L}
&=&\frac{1}{1-\beta\zeta}
\end{IEEEeqnarray}
where 
$\beta=\lim_{N\rightarrow\infty}\frac{L }{K}
$,
$\zeta=\lim_{N\rightarrow\infty}\frac{K }{N_{\rm t}}
$ and 
$\xi^2=\lim_{N\rightarrow\infty} \XMe_i^2
$.
Note that the limit in (\ref{e: limit Nt L}) does not converge for $\beta=\zeta=1$. Thus, we limit the analysis to the case that $\beta\zeta<1$. In the following we will show that the performance  significantly degrade as $\beta\zeta$ approaches $1$, and hence the limiting case $\beta
\zeta\rightarrow 1$ is not interesting. 
Assuming that the bound $\Phi \triangleq\lim_{\substack{N\rightarrow\infty}}\frac{K}{(1-\XMe_i^2)N_{\rm t} }$ exists, and substituting together with (\ref{e: S bound total})-(\ref{e: limit Nt L}) into (\ref{e: uesr rate total for balancing}) gives
\begin{IEEEeqnarray}{rCl}\label{e: uesr rate total limit}
\lim_{N \rightarrow \infty}R_i
\TwoOneColumnAlternate{&\hspace{-1mm}\ge\hspace{-1mm}&}{&\ge&}
E\left[
\log_2\left(1+\frac{\rho}{\frac{1}{1-\beta\zeta}\Phi\cdot\left(\rho(1-\beta (1-\xi^2)) +1\right)}\right)
\right] .
\TwoOneColumnAlternate{\nonumber \\}{}
\end{IEEEeqnarray}

We next consider the two cases of the theorem: If $N_{\rm t}$ grows faster than a logarithmic function of $N$, we have $\zeta=0$, 
\begin{IEEEeqnarray}{rCl}
\xi^2=\lim_{N\rightarrow\infty} 2^{-\frac{c_\mathrm{f} T\log_2 N}{N_{\rm t} }}=1
\end{IEEEeqnarray}
and
\begin{IEEEeqnarray}{rCl}
\Phi&=&\lim_{\substack{N\rightarrow\infty}}\frac{K}{N_{\rm t}}\cdot \frac{1}{ 1-2^{-\frac{F_i T}{N_{\rm t} }} }
\nonumber \\
&=&\lim_{\substack{\alpha \rightarrow 0}}\alpha\cdot \frac{1}{ 1-2^{-\frac{c_\mathrm{f} T }{c_{\rm u} }\alpha} }
\nonumber \\
&=&\frac{c_\mathrm{u}\log_2 e}{c_\mathrm{f} T  }.
\end{IEEEeqnarray}
where we substituted $\alpha=K/N_{\rm t}$. Thus, (\ref{e: uesr rate total limit}) simplifies into (\ref{e: final uesr rate total}). On the other hand, if $N_\mathrm{t}=c_\mathrm{t} \log_2 N$, we get $\zeta=c_\mathrm{u}/c_\mathrm{t}$, $\xi^2= 2^{-\frac{c_\mathrm{f} T}{c_{\rm t} }}$, $\Phi=\frac{c_{\rm u}/c_{\rm t}}{ 1-2^{-\frac{c_{\rm f} T}{c_{\rm t} }} }$ and
\begin{IEEEeqnarray}{rCl}
\TwoOneColumnAlternate{&&}{}
\lim_{N \rightarrow \infty}R_i
\TwoOneColumnAlternate{\\ \nonumber&& \hspace{9mm}\ge}{&\ge&}
E\left[
\log_2\left(1+\frac{\rho}{\frac{1}{1-\beta c_\mathrm{u}/c_\mathrm{t}}\frac{\frac{c_{\rm u}}{c_{\rm t}}\cdot\left(\rho(1-\beta (1-2^{-\frac{c_\mathrm{f} T}{c_{\rm t} }})) +1\right)}{ 1-2^{-\frac{c_{\rm f} T}{c_{\rm t} }} }}\right)
\right] 
\end{IEEEeqnarray}
which is easily simplified to (\ref{e: uesr rate limit ct}) with $\Delta=0$. 

The reader may note that the proof is more general than the theorem, as it allows the use of partial ZF for any number of users between $0$ and $K-1$, while the theorem only addresses the two extreme cases ($\beta=0$ or $\beta=1$). Interestingly, it is easy to show that partial ZF cannot lead to any advantage in the analyzed scenario, and will always be inferior to either the BF or the ZF scheme. To verify this point, one can test the derivative of (\ref{e: uesr rate total limit}) with respect to $\beta$. The sign of this derivative satisfies:
 \begin{IEEEeqnarray}{rCl}\label{e: derivative sign}
\sgn\left\{\frac{d \lim_{N \rightarrow \infty}R_i}{d \beta}\right\}=\sgn\left\{\rho(1-\xi^2)-\zeta \left(1+\rho\right)\right\}
\IEEEeqnarraynumspace\end{IEEEeqnarray}
which does not depend on $\beta$. Thus, the optimal value of $\beta$ can only be one of its extreme values: $0$ (corresponding to BF) or $1$ (corresponding to ZF). 

\subsubsection{Proof for DP}

To complete the proof of Theorem \ref{Th: Scaling}, we next present and analyze the DP precoding scheme. More specifically, we consider the dirty paper scheme with zero forcing linear pre-processing (e.g., \cite{ginis2002vectored}). We use the LQ decomposition of the quantized channel matrix:
\begin{IEEEeqnarray}{rCl}\label{e: LQ decomp}
\hHm=\Lm \Qm
\end{IEEEeqnarray}
where $\Lm$ is a lower triangular matrix and $\Qm$ is a unitary matrix, and
apply a precoding of the data symbols:
\begin{IEEEeqnarray}{rCl}
\xv=\Qm^H \begin{bmatrix}\sv\\ \bf 0\end{bmatrix}
\end{IEEEeqnarray}
where $\xv$ is the vector of transmitted symbols and $\sv$ is the $K \times 1$ vector that contains the actual data symbols. Note that the precoding does not change the power of the resulting transmitted vector. Thus, using $E[\sv\sv^H]=\frac{1}{K}\mathbf{I}$ gives  $E[\xv^H\xv]=E[\sv^H\sv]=1$. The resulting effective channel is:
\begin{IEEEeqnarray}{rCl}\label{e: revised channel model}
\yv =  \sqrt{\rho}\Lm \sv + \sqrt{\rho}\eM\xv +\nv.
\end{IEEEeqnarray}
Recalling that $\Lm$ and $\sv$ are known to the BSs while  $\eM$ is not known,
we define:
\begin{IEEEeqnarray}{rCl}
z_i=\sum_{j=1}^{i-1} \ell_{i,j}s_j
\end{IEEEeqnarray}
\begin{IEEEeqnarray}{rCl}\label{e: effective noise term}
w_i=n_i+ \sqrt{\rho}\sum_{j=1}^{N_\textrm{t}}  \eMe_{i,j}x_j 
\end{IEEEeqnarray}
and we can write:
\begin{IEEEeqnarray}{rCl}\label{e: Lemma channel model}
y_i =  \sqrt{\rho}\ell_{i,i} s_i +z_i+w_i .
\end{IEEEeqnarray}
where $ \ell  _{i,i}$ is the $i$-th element on the diagonal of the matrix $\Lm$.
The zero mean of $\eMe_{i,j}$ and its statistical independence on $\hHm$ guarantee that $E[s_i w_i]=E[z _i w_i]=0$. Hence, 
we use the following lemma to lower bound the rate.
\begin{lemma}\label{Lem: lower bound simple}
In the channel model of (\ref{e: Lemma channel model}), if $E[s_i w_i]=0$ then the channel capacity is lower bounded by:
\begin{IEEEeqnarray}{rCl}\label{e: lemma 1 eq}
R
&\ge&
E\left[
\log_2\left(1+\frac{\rho|\ell_{i,i} |^2 E[|s_i|^2]}{E[|w_i|^2]}\right)
\right] .
\end{IEEEeqnarray}
\end{lemma}
\begin{IEEEproof}
Lemma \ref{Lem: lower bound simple} can be directly proved using Theorem 1 of \cite{Bergel2014Letter}, which guarantees the applicability of the dirty paper scheme even if the noise, $w_i$ is not Gaussian, and possibly even statistically dependent on $s_i$ and $z_i$. An alternative proof of this theorem, \cite{Bergel2014SPAWC}, also showed that bound can be achieved using a multidimensional lattice precoding. 
\end{IEEEproof}

Thus we have: 
\begin{IEEEeqnarray}{rCl}\label{e: Ri from full version}
R_i
&\ge& E\left[\log_2\left(1+\frac{\rho|\ell_{i,i} |^2 E[|s_i|^2]}{1+\rho  \XMe_i^2 E[\xv^H\xv]}\right)\right]
\nonumber \\
&=&E\left[\log_2\left(1+\frac{\rho|\ell_{i,i} |^2 }{K+\rho K \XMe_i^2 }\right)\right].
\end{IEEEeqnarray}
where we used 
$E[|s_i|^2]=1/K$. Using Lemma 2.1 of \cite{tulino2004random} (originally from \cite{gupta1999matrix}) $2 |\ell_{i,i}|^2/(1-\XMe_i^2)$ has a Chi square distribution with $2(N_\textrm{t}-i+1)$ degrees of freedom. 

To homogenize the system we use random permutations between the users. Thus, we get identical rates for all users. Denoting the rates before the randomization as $R^\mathrm{o}_i$, and averaging the achievable rate over all users, gives:
\begin{IEEEeqnarray}{rCl}\label{e: Ri from full version 1}
R_i&=&\lim_{N_\textrm{t}\rightarrow \infty}\frac{1}{K}\sum_{k=1}^K R^\mathrm{o}_k
\nonumber \\&\ge& 
\lim_{\substack{N_\textrm{t}\rightarrow \infty \\ k=\lfloor \alpha K\rfloor}}\int_0^1E\left[\log_2\left(1+\frac{N_\textrm{t}(1-\XMe_k^2)}{K}\frac{\rho\frac{|\ell_{k,k} |^2}{N_\textrm{t}(1-\XMe_k^2)} }{1+\rho  \XMe_k^2 }\right)\right]d\alpha
\nonumber \\
&=&
\int_0^1 \log_2\left(1+\frac{1}{\Phi}\frac{\rho(1-\alpha\zeta) }{1+\rho  \xi^2 }\right)d\alpha 
\end{IEEEeqnarray}
where we used again $\zeta=\lim_{N_\textrm{t}\rightarrow \infty}\frac{ K}{N_\textrm{t}}$, $\xi^2=\lim_{N\rightarrow\infty} \XMe_i^2
$
 and $\Phi \triangleq\lim_{\substack{N\rightarrow\infty}}\frac{K}{(1-\XMe_i^2)N_{\rm t} }$
, and also substituted the bound of the diagonal elements of the matrix $\Lm$ for $0\le \alpha\le 1$:
\begin{IEEEeqnarray}{rCl}
\lim_{\substack{N_\textrm{t}-K\rightarrow \infty\\ k=\lfloor \alpha K\rfloor}}  \frac{|\ell_{k,k} |^2}{N_\textrm{t}(1-\XMe_i^2)} =\lim_{N_\textrm{t}\rightarrow \infty}  \frac{N_\textrm{t}-\alpha K+1}{N_\textrm{t}}=1-\alpha\zeta .
\IEEEeqnarraynumspace\end{IEEEeqnarray}

In the first case of the theorem,  $N_{\rm t}$ grows faster than a logarithmic function of $N$. From Sub-subsection  \ref{subsub: linear precoding scaling} we have $\zeta=0$, $\xi^2=1$
and
$
\Phi=\frac{c_\mathrm{u}\log_2 e}{c_\mathrm{f} T  }$. In this case, the integral trivially simplifies to (\ref{e: final uesr rate total}). 

For the second case, we have $N_\mathrm{t}=c_\mathrm{t} \log_2 N$, which leads to $\zeta=c_\mathrm{u}/c_\mathrm{t}$, $\xi^2= 2^{-\frac{c_\mathrm{f} T}{c_{\rm t} }}$ and $\Phi=\frac{c_{\rm u}/c_{\rm t}}{ 1-2^{-\frac{c_{\rm f} T}{c_{\rm t} }} }$. Thus, (\ref{e: Ri from full version 1})
becomes:
\begin{IEEEeqnarray}{rCl}
R_i
&\ge&
\int_0^1 \log_2\left(1+\frac{ 1-2^{-\frac{c_{\rm f} T}{c_{\rm t} }} }{c_{\rm u}/c_{\rm t}}\frac{\rho(1-\alpha\frac{c_\mathrm{u}}{c_\mathrm{t}}) }{1+ \rho2^{-\frac{c_\mathrm{f} T}{c_{\rm t} }} }\right)d\alpha .
\end{IEEEeqnarray}

Substituting $u=(1-\alpha\frac{c_\mathrm{u}}{c_{\rm t}})$ and $\gamma =\rho\frac{1-2^{- c_\mathrm{f}  T / c_\mathrm{t}  }}{1+\rho  2^{- c_\mathrm{f}  T / c_\mathrm{t}  }}$ gives:
\begin{IEEEeqnarray}{rCl}\label{e: another equation in the DP proof}
 R_i
&\ge&\frac{c_\mathrm{t}}{c_{\rm u}}
\int_{1-\frac{c_\mathrm{u}}{c_{\rm t}}}^1\log_2\left(1+\frac{c_\mathrm{t}}{c_{\rm u}}\gamma u\right)d u
\nonumber \\
&=&\frac{c_\mathrm{t}}{c_{\rm u}} \left.\frac{-\frac{c_\mathrm{t}}{c_{\rm u}}\gamma  u \log_2(e) + (1 + \frac{c_\mathrm{t}}{c_{\rm u}}\gamma  u) \log_2(1 + \frac{c_\mathrm{t}}{c_{\rm u}}\gamma  u)}{\frac{c_\mathrm{t}}{c_{\rm u}}\gamma }\right|_{1-\frac{c_\mathrm{u}}{c_{\rm t}}}^1
\nonumber \\
&=& \frac{ (1 + \frac{c_\mathrm{t}}{c_{\rm u}}\gamma ) \log_2(1 +\frac{c_\mathrm{t}}{c_{\rm u}} \gamma  )}{\gamma }-\log_2(e)
\nonumber \\
&&- \frac{ (1 + \frac{c_\mathrm{t}}{c_{\rm u}}\gamma ( 1-\frac{c_\mathrm{u}}{c_{\rm t}})) \log_2(1 + \frac{c_\mathrm{t}}{c_{\rm u}}\gamma (1-\frac{c_\mathrm{u}}{c_{\rm t}}))}{\gamma } .
\end{IEEEeqnarray}
Adding the term $(1 + \frac{c_\mathrm{t}}{c_{\rm u}}\gamma ) \log_2(1 + \frac{c_\mathrm{t}}{c_{\rm u}}\gamma (1-\frac{c_\mathrm{u}}{c_{\rm t}}))$
in the last line of (\ref{e: another equation in the DP proof}) and subtracting it from the previous line simplifies the expression to:
\begin{IEEEeqnarray}{rCl}
R_i
&\ge&
-\frac{ (1 + \frac{c_\mathrm{t}}{c_{\rm u}}\gamma ) \log_2\left(\frac{1 + \frac{c_\mathrm{t}}{c_{\rm u}}\gamma (1-\frac{c_\mathrm{u}}{c_{\rm t}})}{1 +\frac{c_\mathrm{t}}{c_{\rm u}} \gamma }\right )}{\gamma }-\log_2(e)
\nonumber \\
&&+\log_2(1 + \frac{c_\mathrm{t}}{c_{\rm u}}\gamma (1-\frac{c_\mathrm{u}}{c_{\rm t}}))
\end{IEEEeqnarray}
which simplifies to (\ref{e: uesr rate limit ct}) and (\ref{d: Delta DP}) with $\beta=1$.

\TwoOneColumnAlternate{
\FigDat{Balancing}{0.5}
}{
\FigDat{Balancing}{0.8}
}

%%%%%%%%%%%%%%%%%%%%%%%%%%%%%%%%%%%%
\section{Numerical examples}\label{sec: numeric}
\TwoOneColumnAlternate{
In this section we demonstrate the results derived above, by presenting simulation results for a massive MIMO system. 
Fig. \ref{f:Balancing fig} demonstrates the rate balancing principle of Theorem \ref{Th: DP balancing}. The figure depicts the average user rate versus the number of feedback bits sent in each coherence block (i.e., $F_i T$). The average rate is depicted for systems that perform beamforming (BF) zero-forcing (ZF) and dirty paper coding (DP), all with $N_\mathrm{t}=8$ transmit antennas and $K=8$ users.
As expected, the BF scheme give reasonable performance for low feedback rate, but can't gain much from high feedback rate because it is interference limited. On the other hand, both the ZF and DP scheme gain significantly from the increase in feedback rate, and the average user rate goes up nearly linearly. Furthermore, one can easily verify that the slope of the rate matches the $1/N_\mathrm{t}$ slope predicted in  (\ref{e: th1_bound}).

\FigDat{Performance}{0.5}
}{
\FigDat{Performance}{0.8}
}

Obviously, the linear increase of the rate with the feedback can only happen in an interference limited system. Thus, the figure shows that for SNR of $\rho=30$dB the spectral efficiency flattens out around 6 and 8 Bps/Hz in the ZF and DP schemes, respectively. Yet, such a spectral efficiency is much better than what we have in current state of the art systems with the same SNR. 

To demonstrate the scaling of the rate with the number of antennas, Fig. \ref{f:Performance fig} depicts the average spectral efficiency per user at the downlink as a function of the number of transmit antennas. For simplicity, this figure depicts only the performance of the BF system for the case $c_\mathrm{u}=4$,
$c_\mathrm{f}T=10$, i.e., the number of users is set to the integer that is closest to $4\cdot \log_2(N)$ and the number of feedback bits takes $\frac{10}{T}$ of the uplink rate of each user (e.g., if the coherence block contains $180$ symbols as in the example of \cite{bergel2014balancing_submitted}, then approximately $5.6\%$ of the uplink is devoted to feedback).

The figure shows three different curves for different choices of the number of transmit antennas. The $2$ lower curves correspond to logarithmic scaling of the number of transmit antennas with the number of receive antennas, while the highest curve uses all of the available antennas for transmission. In all cases, the simulated spectral efficiency (per user) converges to the asymptotic results of Theorem \ref{Th: DP balancing} (when the number of antennas is large enough). Thus, the system sum rate scales logarithmically with the number of antennas. This logarithmic scaling is much slower than the linear scaling of the uplink. Yet, it can still give significant throughput gains. As an example, the figure give more details for the case of $4$ and $128$ antennas (note that the number of transmit antennas is upper bounded by the total number of antennas). A comparison of these two points shows that increasing the number of antennas from $4$ to $128$ results in a multiplication of the sum-rate by more than $5$ for all three setups. 

\TwoOneColumnAlternate{
\FigDat{TotalPerformance}{0.5}
}{
\FigDat{TotalPerformance}{0.8}
}

To further demonstrate the massive MIMO gain, Fig. \ref{f:TotalPerformance fig} depicts the sum of all user rates as a function of the number of receive antennas for the three analyzed schemes. As can be seen, in all cases we have a logarithmic growth of the system throughput with the number of antennas. Furthermore, as predicted by (\ref{e: final uesr rate total}), when all antennas are used for transmission ($N_\mathrm{t}=N$) all schemes converge to the same performance. As stated above, it is more practical to limit the number of transmit antennas. Hence, it is good to observe that logarithmic scaling is also obtained using a logarithmic number of antennas. Note that all antennas are required to be used for reception. This is in order to maintain the logarithmic growth of the uplink rate with the number of antennas, which in turn is needed to support the required feedback rate. 

%%%%%%%%%%%%%%%%%%%%%%%%%%%%%%%%%%%%
\section{conclusions}\label{sec: discussion}
This work studied the extension of the uplink-downlink balancing principle to FDD massive MIMO systems. 
The uplink-downlink balancing concept is based on using part of the uplink capacity for CSI feedback in order to increase the downlink capacity. We extended the proof of the balancing principle, to the case of a massive MIMO system that performs linear zero-forcing, showing that any unused uplink rate can be converted to an increase in the downlink rate with a favorable exchange ratio. The possibility to tradeoff uplink rate for downlink rate is very  important in FDD massive MIMO systems, as the use of many receive antennas yields significant gains for the uplink throughput.

We further study the case  in which a constant fraction of the uplink rate is devoted to CSI feedback, and show that the downlink sum-rate  can scale logarithmically with the number of antennas. This scaling is achieved by scaling the number users logarithmically with the number of transmit antennas and achieving a constant throughput for each user. While this scaling is lower than the uplink rate scaling or the TDD rate scaling, it still brings good throughput gains. 

The practicality of the suggested approach is further emphasized by showing that the logarithmic scaling is achievable even when only using a small fraction of the antennas for transmission. More specifically, the number of transmit antennas is required to only scale logarithmically with the number of receive antennas. This limits the number of downlink pilot symbols needed, which significantly improves the system efficiency.   

\section{Acknowledgement}
We wish to thank Dr. Shimon Moshavi for his valuable advice and
comments that helped us improve this manuscript.

%%%%%%%%%%%%%%%%%%%%%%%%%%%%%%%%%%%%%%%%%%%%
%\begin{appendices}
%%%%%%%%%%%%%%%%%%%%%%%%%%%%%%%%%%%%
%\end{appendices}

\bibliographystyle{IEEEtran}
\bibliography{../../Bergel_2015_all_bib}
%\bibliography{ULDL_Jul_18}
\clearpage

%\TwoOneColumnAlternate{}{\FigDat{System}{0.9}\FigDat{Performance}{0.9}}

\end{document}